\documentstyle[prb,twocolumn,aps,epsf]{revtex}
\begin{document} \draft
\title{Bifurcation at the $c=3/2$ Takhtajan-Babujian point \\
to the $c=1$ critical lines}
\author{Atsuhiro Kitazawa and Kiyohide Nomura}
\address{Department of Physics, Kyushu University, Fukuoka 812-8581, Japan}
\date{\today}
\maketitle
\begin{abstract}
We study the $S=1$ quantum spin chains with bilinear, biquadratic, plus 
bond alternation in the vicinity of the $S=1$ Takhtajan-Babujian model. 
Transition line between the Haldane and the dimer phases 
are determined numerically. 
To see the crossover behavior from $c=3/2$ ($k=2$ SU(2) WZW model) 
at the Takhtajan-Babujian point to $c=1$ ($k=1$ SU(2) WZW model), 
we calculate the conformal anomaly $c$ 
and scaling dimensions of the primary fields on the transition line. 
\end{abstract}
\pacs{75.10.Jm,11.10.Hi,64.60.Fr,75.40.Cx}
\section{Introduction}
Critical phenomenon is one of the most important topics in physics. 
For the two dimensional classical and the one dimensional quantum 
systems, the conformal invariance plays an crucial role for the 
understanding and the classification of it\cite{cft}. 
Then the next question becomes important: 
how does the crossover occur between different fixed points? 
Zamolodchikov \cite{Zamolodchikov} argued for this problem in general 
situation, and derived `c-theorem'. 
This theorem says that if there exists a renormalization group flow 
from one (ultraviolet) fixed point to another (infrared) fixed point, 
the conformal anomaly $c$ should decrease. 
With perturbation expansions, 
Zamolodchikov examined this situation between minimal models 
(see also ref. \cite{Ludwig}). 

On the other hand, 
Affleck and Haldane\cite{Affleck87} studied quantum spin chains 
based on the SU(2) Wess-Zumino-Witten (WZW) model 
using non-Abelian bosonization and O(3) $\sigma$ model. 
They argued the instability of the integrable Takhtajan-Babujian (TB) 
model\cite{Takhtajan,Babujian} 
which is described by the $k=2S$ SU(2) WZW model\cite{Affleck86,Affleck86b}. 
Introduction of relevant couplings by the bond alternation
makes the $k=2S$ system unstable 
and develops the excitation gap, but some combinations of the couplings 
reduce the central charge $k$ (or the topological coupling of the WZW model) 
to $1$. 
In this situation, the conformal anomaly decreases from $c=3k/(k+2)$ to $c=1$, 
which is consistent to the Zamolodchikov's $c$-theorem. 
And the TB model can be a multi-critical point\cite{Affleck87}. 

The crossover behavior from $k=3$ to $k=1$ SU(2) WZW model 
was investigated by Ziman and Schulz\cite{Ziman} 
with the numerical diagonalization for the $S=3/2$ model. 
They showed the behavior from the $S=3/2$ TB model to the Heisenberg model 
(with next nearest neighbor interactions). 

In this paper we discuss the other type crossover behavior, that is, 
from $k=2$ to $k=1$ SU(2) WZW model. 
To see this behavior, we study the following 
$S=1$ spin model in the vicinity of the TB point 
($\theta = -\pi/4$, $\delta=0$), 
\begin{equation}
  H = \sum_{j=1}^{N} \left(1-\delta(-1)^{j}\right)\left[
  \cos\theta\mbox{\boldmath$S$}_{j}\cdot \mbox{\boldmath$S$}_{j+1}
  +\sin\theta(\mbox{\boldmath$S$}_{j}\cdot \mbox{\boldmath$S$}_{j+1})^{2}
  \right]. 
\label{hamiltonian}
\end{equation}
For $\delta=0$ several authors studied 
numerically,\cite{Solyom,Oitmaa,Blotea,Chang,Fath} 
and nowaday it is widely believed that 
a transition occurs at $\theta = -\pi/4$ 
between the $S=1$ Haldane ($\theta>-\pi/4$) 
and the dimer phases ($\theta<-\pi/4$). 
The ground state of the former phase 
is singlet, but for the latter phase, the ground state is two fold degenerate. 
In the Haldane phase, the ground state 
at the point $\theta=\mbox{arctan}(1/3)$ is known rigorously.\cite{AKLT} 
This phase is characterized by the string type 
order parameter\cite{Nijs,Tasaki}
and the four fold degeneracy of the ground state 
for the open boundary condition.\cite{AKLT,Kennedy90} 
These are related to a hidden $Z_{2}\times Z_{2}$ symmetry.\cite{Kennedy92} 
In the dimer phase, the one-site translational invariance is 
broken spontaneously, and neighboring two spins make a singlet. 
Near $\theta=-\pi/4$, the energy gap\cite{Affleck86,Fath} behaves 
as $|\theta +\pi/4|$. 

With the bond alternation $\delta$, 
the non-degenerate dimer ground state appears, 
and in the region $-3\pi/4<\theta<-\pi/4$, the first order phase transition 
occurs at $\delta=0$.\cite{Affleck87,Solyom,blote87} 
There exists the transition line between the Haldane 
and the dimer phases. 
In this case, the critical properties are of the $k=1$ SU(2) WZW theory. 
This behavior has been studied for $\theta=0$, $\delta\approx 0.26$ 
extensively.\cite{Kato,Yamamoto,Totsuka,KN97b} 

In this paper, we study the model (\ref{hamiltonian}) 
with the numerical diagonalization for finite size systems, 
to see the critical behaviors on the $c=1$ critical line near the TB point. 
Our obtained phase diagram is shown in Fig. 1. 
There exist two $c=1$ critical lines which separate the $S=1$ Haldane 
and two dimer phases. 
At the TB point, these two lines meet a first order phase transition line 
($\theta<-\pi/4$, $\delta=0$), on which the ground states are 
two-fold degenerate and have an excitation gap. 
The TB point is an multi-critical point and the critical behavior is described 
by the $k=2$ SU(2) WZW model with $c=3/2$. 
On the other termination point of the $c=1$ line 
$(\theta,\delta) = (\mbox{arctan}(1/3),\pm 1)$, there are highly degenerate 
ground states. 

The organization of this paper is as follows. 
In the next section, to see the instability of the $c=3/2$ system, 
we consider the model based on 
the three component Majorana fermions as an effective one. 
We identify primary fields of the $k=2$ SU(2) WZW model with of 
the fermion model in SU(2) irreducible forms, 
and give some renormalization group arguments. 
In section III, using twisted boundary condition method 
we determine the $c=1$ critical line 
with numerical diagonalization for finite size systems ($N=8,10,12,14,16$). 
To see the crossover behavior from $c=3/2$ to $c=1$ systems, 
we show the conformal anomaly 
and scaling dimensions of the primary fields on the critical line. 
Since we can only treat finite size systems in numerical diagonalization, 
we analyze the finite size behavior of conformal anomaly in sec. II 
and compare it with the numerical results in subsection IIIB. 
The last section gives summary and discussion. 

\section{Real fermions}
At the TB point ($\theta=-\pi/4$, $\delta=0$), 
the critical properties of the system are 
described by the $k=2$ SU(2) Wess-Zumino-Witten model.\cite{Affleck86} 
This has been confirmed by Bethe ansatz.\cite{Alcaraz88,Frahm,Klumper} 
Near this point, we have the following continuum effective Hamiltonian 
\begin{eqnarray}
  H &=& \int\frac{dx}{2\pi} \frac{1}{4}
  \left[ :\mbox{\boldmath$J$}(x)\cdot\mbox{\boldmath$J$}(x): + 
  :\bar{\mbox{\boldmath$J$}}(x)\cdot\bar{\mbox{\boldmath$J$}}(x): \right]
  \nonumber \\
  &&  -\frac{g_{1}}{2\pi}\int dx \Phi_{1/2,1/2}^{0,0}(x)
      -\frac{g_{2}}{2\pi}\int dx \Phi_{1,1}^{0,0}(x)
\label{WZW}  
\\
  &&  -\frac{g_{0}}{2\pi}\int dx \frac{1}{\sqrt{3}}
  \mbox{\boldmath$J$}(x)\cdot\bar{\mbox{\boldmath$J$}}(x)
\nonumber
\end{eqnarray}
where $\mbox{\boldmath$J$}$ and $\bar{\mbox{\boldmath$J$}}$ 
are the SU(2) currents for the left and right movers, 
the symbol ``$: :$'' means the normal ordering. 
The operator $\Phi_{s,\bar{s}}^{S,S^{z}}$ is one of 
the primary fields whose conformal weight 
is given by $(s(s+1)/4,\bar{s}(\bar{s}+1)/4)$, 
where $s$ and $\bar{s}$ are the spin of the left and the right movers, 
and $S$ and $S^{z}$ are the total spin and the total third component of them. 
The last term gives a marginal correction. 
At the TB point the couplings $g_{1}$ and $g_{2}$ vanish, and $g_{0}>0$ 
(so that the last term is marginally irrelevant, see eq.(\ref{RGeq})). 
The correspondence to the $S=1$ spin model (\ref{hamiltonian}) is as follows
\begin{equation}
  g_{1}\propto\delta, \hspace{5mm} g_{2}\propto \theta+\frac{\pi}{4}. 
\end{equation}

For the one-site translational invariant case in the spin chain 
($\delta=0$), $g_{1}$ vanishes, 
and the transition between the two-fold degenerate dimer and the Haldane 
phases is induced by 
the third term of (\ref{WZW}) ($\Phi_{1,1}^{0,0}$). 
Scaling dimension of the operator $\Phi_{1,1}^{0,0}$ is $x_{1,1}^{0}=1$, and 
near the TB point the excitation gap behavior is 
$\Delta E \sim |g_{2}| \sim |\theta+\pi/4|$ 
as is the 2D Ising transition.\cite{Affleck86}  

On the other hand, the $k=2$ SU(2) WZW model can also be described by 
the three component Majorana fermions,\cite{Zamolodchikov86} 
\begin{equation}
H_{0} = \frac{1}{2}\int dx \sum_{a=1}^{3}\left( i\psi_{a}\frac{d}{dx}\psi_{a} 
        - i\bar{\psi}_{a}\frac{d}{dx}\bar{\psi}_{a}\right),
\label{Majorana}
\end{equation}
where the fermions satisfy the anti-commutation relation 
$\{\psi_{a}(x),\psi_{b}(x^{'})\}=\delta_{ab}\delta(x-x{'})$, 
and the SU(2) current can be written as 
\begin{eqnarray}
  J^{a}(x) &=& -\frac{2\pi}{2} i\epsilon_{abc}\psi_{b}(x)\psi_{c}(x), 
\nonumber \\
  \bar{J}^{a}(x) &=& -\frac{2\pi}{2}i\epsilon_{abc}
      \bar{\psi}_{b}(x)\bar{\psi}_{c}(x). 
\end{eqnarray}
The correlation functions of the fermions are given by 
\begin{eqnarray}
  \langle T_{\tau}[\psi_{a}(\tau,x)\psi_{b}(\tau^{'},x^{'})]\rangle
  &=& \frac{\delta_{ab}}{2\pi}\frac{1}{\tau-\tau^{'} + i (x - x^{'})},
  \nonumber \\
  \langle T_{\tau}[\bar{\psi}_{a}(\tau,x)\bar{\psi}_{b}(\tau^{'},x^{'})]\rangle
  &=& \frac{\delta_{ab}}{2\pi}\frac{1}{\tau-\tau^{'} - i (x - x^{'})},
\end{eqnarray}
where $T_{\tau}$ is the imaginary time ordering. 
The operator product expansions (OPEs) of the current fields are 
\begin{eqnarray}
  J^{a}(z)J^{b}(0) &=& \frac{\delta_{ab}}{z^{2}} 
  + \frac{i\epsilon_{abc}}{z}J^{c}(z) + \cdots, \nonumber \\
  \bar{J}^{a}(\bar{z})\bar{J}^{b}(0) &=& \frac{\delta_{ab}}{\bar{z}^{2}} 
  + \frac{i\epsilon_{abc}}{\bar{z}}\bar{J}^{c}(\bar{z}) + \cdots,
\end{eqnarray}
where $z=\tau+ix$ and $\bar{z}=\tau-ix$. 
Then the primary fields with the conformal weight $(1/2,1/2)$ are written by 
the SU(2) irreducible operators $\Phi^{S,S^{z}}_{1,1}$ as 
\begin{eqnarray}
  \Phi^{2,2}_{1,1}(x) &=& 
  \pi i\left(\psi_{1}(x) + i\psi_{2}(x)\right)
  \left(\bar{\psi}_{1}(x) + i\bar{\psi}_{2}(x)\right),
  \nonumber \\
  \Phi^{2,1}_{1,1}(x) &=& 
  -\pi i\left(\psi_{1}(x)+i\psi_{2}(x)\right)\bar{\psi}_{3}(x)
\nonumber \\
  && -\pi i\psi_{3}(x)\left(\bar{\psi}_{1}(x)+i\bar{\psi}_{2}(x)\right),
  \nonumber \\
  \Phi^{2,0}_{1,1}(x) &=&
    \frac{2\pi i}{\sqrt{6}}\left( 2\psi_{3}(x)\bar{\psi}_{3}(x) \right.
\nonumber \\
  && \left.  - \psi_{1}(x)\bar{\psi}_{1}(x) 
     - \psi_{2}(x)\bar{\psi}_{2}(x) \right),
  \nonumber \\
  \Phi^{1,1}_{1,1}(x) &=& 
  \pi i\left(\psi_{1}(x)+i\psi_{2}(x)\right)\bar{\psi}_{3}(x)
  \\
  &&-\pi i\psi_{3}(x)\left(\bar{\psi}_{1}(x)+i\bar{\psi}_{2}(x)\right),
\nonumber \\
  \Phi^{1,0}_{1,1}(x) &=&
  \sqrt{2}\pi i\left(
  \psi_{1}(x)\bar{\psi}_{2}(x) - \psi_{2}(x)\bar{\psi}_{1}(x)\right), 
  \nonumber \\
  \Phi^{0,0}_{1,1}(x) &=&
  \frac{2\pi i}{\sqrt{3}}\left( \psi_{1}(x)\bar{\psi}_{1}(x)
  + \psi_{2}(x)\bar{\psi}_{2}(x) \right. \nonumber \\
  && \left. + \psi_{3}(x)\bar{\psi}_{3}(x) \right),
  \nonumber
\end{eqnarray}
where we only show the operators with $S^{z}\geq0$ (operators with $S^{z}<0$ 
are the Hermitian conjugate of the above ones). 
This fermion model was considered by Tsvelik\cite{Tsvelik} 
to study the magnetic behavior of the $S=1$ anisotropic spin chains, 
and Shelton {\it et al}.\cite{Shelton} derived the model 
from the spin ladder system. 

The Hamiltonian (\ref{Majorana}) is a simple sum of the three independent 
Majorana fermion models. 
By the way, one-component model describes the 2D Ising transition. 
Then the operator $\Phi^{0,0}_{1,1}$ is 
the sum of energy densities ($\epsilon$). 
The order ($\sigma$) and the disorder ($\mu$) parameters 
are primary fields of the $c=1/2$ CFT, 
and one can write down the primary fields $\Phi^{S,S^{z}}_{1/2,1/2}$ 
with this operators. 
Using the operator product expansion\cite{Francesco}
\begin{eqnarray}
  \lefteqn{\sigma(z,\bar{z})\mu(0,0)} \nonumber \\
  && = \frac{1}{\sqrt{2}|z|^{1/4}}\left(
  e^{i\pi/4}z^{1/2}\psi^{'}(0) 
  + e^{-i\pi/4}\bar{z}^{1/2}\bar{\psi^{'}}(0)\right) + \cdots,
\end{eqnarray}
(where we redefined $\psi^{'}_{a} = \sqrt{2\pi}\psi_{a}$) 
the SU(2) irreducible operators with the conformal 
weight $(h, \bar{h})=(3/16,3/16)$ are following 
two types,\cite{Zamolodchikov86,Tsvelik,Shelton} 
\begin{eqnarray}
  \Phi^{1,1}_{1/2,1/2} &=& \frac{1}{\sqrt{2}}
    (\sigma_{1}\mu_{2}+i\mu_{1}\sigma_{2})\mu_{3},
  \nonumber \\
  \Phi^{1,0}_{1/2,1/2} &=& \mu_{1}\mu_{2}\sigma_{3},
  \\
  \Phi^{0,0}_{1/2,1/2} &=& \sigma_{1}\sigma_{2}\sigma_{3},
  \nonumber
\end{eqnarray}
and 
\begin{eqnarray}
  \Phi^{'1,1}_{1/2,1/2} &=& \frac{1}{\sqrt{2}}
    (\mu_{1}\sigma_{2}+i\sigma_{1}\mu_{2})\sigma_{3},
  \nonumber \\
  \Phi^{'1,0}_{1/2,1/2} &=& \sigma_{1}\sigma_{2}\mu_{3},
  \\
  \Phi^{'0,0}_{1/2,1/2} &=& \mu_{1}\mu_{2}\mu_{3}.
  \nonumber
\end{eqnarray}
These two representations are distinguished by the parity 
of the number of order parameters. 
Here we choose the first one, so that $g_{2}>0$ for $\theta < -\pi/4$. 

After the calculation of the OPEs among 
$\Phi_{1/2,1/2}^{0,0}$, $\Phi_{1,1}^{0,0}$, 
and $3^{-1/2}\mbox{\boldmath$J$}\cdot \bar{\mbox{\boldmath$J$}}$, 
we have the following one-loop renormalization group equations, 
\begin{eqnarray}
\frac{dg_{1}(L)}{d\ln L} &=& \left(2-\frac{3}{8}\right)g_{1}(L)
    +\frac{\sqrt{3}}{2}g_{1}(L)g_{2}(L) \nonumber \\
    && -\frac{\sqrt{3}}{4}g_{1}(L)g_{0}(L), \nonumber \\
\frac{dg_{2}(L)}{d\ln L} &=& \left(2-1\right)g_{2}(L)
     +\frac{\sqrt{3}}{4}(g_{1}(L))^{2}
\label{RGeq} \\
    && -\frac{2\sqrt{3}}{3}g_{2}(L)g_{0}(L), \nonumber \\
\frac{dg_{0}(L)}{d\ln L} &=& 
    -\frac{\sqrt{3}}{8}(g_{1}(L))^{2}-\frac{\sqrt{3}}{3}(g_{2}(L))^{2} 
\nonumber \\
    && -\frac{\sqrt{3}}{3}(g_{0}(L))^{2},
\nonumber
\end{eqnarray}
where $L$ is an infrared cut-off. 
These equations are invariant under the transformation 
$g_{1}\rightarrow -g_{1}$, which corresponds to $\delta\rightarrow -\delta$ 
in the spin system (\ref{hamiltonian}). 
(On the other hand, these are not invariant under $g_{2}\rightarrow -g_{2}$ 
for $g_{1}\neq 0$. )
In this order, Zamolodchikov's $c$-function\cite{Zamolodchikov} 
is calculated as
\begin{eqnarray}
  c(g_{1}(L),g_{2}(L)) &=& \frac{3}{2}-\frac{39}{32}(g_{1}(L))^{2}
  -\frac{3}{4}(g_{2}(L))^{2} \nonumber \\
  && -\frac{3\sqrt{3}}{8}(g_{1}(L))^{2}g_{2}(L),
\end{eqnarray}
where we neglect the marginal coupling $g_{0}$, which has 
larger scaling dimension comparing two other operators. 

The RG flow is shown in Fig. 2 where we neglect $g_{0}$. 
It is seen from this figure, that RG flows to two different strong coupling 
regions and there is a second order phase transition line between them. 
We see that near the multi-critical point $g_{1}=g_{2}=0$, 
the second order transition line behaves as 
\begin{equation}
  g_{1c} \sim (g_{2})^{13/8},
\label{eq:line}
\end{equation}
where the exponent $13/8$ is the ratio $(2-x)$ of the two relevant operators. 
For finite $L$, there exists the region that $g_{1}(L)$ and $g_{2}(L)$ is 
small, and that the perturbation theory is applicable. 
In the leading order, the conformal anomaly behaves as 
\begin{eqnarray}
  c(g_{1}(L),g_{2}(L)) &=& \frac{3}{2}-\frac{3}{4}(g_{2}(L))^{2}-\cdots 
  \nonumber \\
    &=& \frac{3}{2}-\alpha (g_{1}(L))^{2(8/13)}-\cdots,
\label{aanmly}
\end{eqnarray}
near the $c=3/2$ point ($\alpha$ is a constant). 
For $g_{1,2}\neq 0$, the couplings $g_{1}(L)$ and $g_{2}(L)$ should flow 
to the $c=1$ fixed point or the $k=1$ SU(2) WZW model. 
In the thermodynamic limit $L\rightarrow \infty$, 
the conformal anomaly should be discontinuous from $c=3/2$ to $c=1$ 
at the TB point. 

\section{$c=1$ model approach and numerical results}
To investigate critical behaviors numerically, it is needed to determine 
the critical point accurately. 
Thus in this section, 
we firstly introduce the method to determine the phase boundary 
between the Haldane and the dimer phases (IIIA). 
In subsection IIIB, 
we show our numerical results, and see the asymptotic 
behaviors of the scaling dimensions and conformal anomaly. 

Assuming the conformal invariance, the scaling dimension $x_{n}$ 
is related to the energy gap of the finite size system with the 
periodic boundary condition\cite{Cardy}
\begin{equation}
  x_{n} = \frac{L}{2\pi v}(E_{n}(L) - E_{g}(L)), 
\label{dimension}
\end{equation}
where $L$ is the system size ($L=N/2$), 
$E_{g}(L)$ is the ground state energy, and $v$ is the sound velocity. 
The leading finite size correction of 
the ground state energy is\cite{Affleck86b,Blote} 
\begin{equation}
  \frac{E_{g}(L)}{L} = \epsilon_{g} - \frac{\pi vc}{6L^{2}},
\label{anomaly}
\end{equation}
where $\epsilon_{g}$ is the ground state energy per site for the infinite 
size system, $c$ is the conformal anomaly. 
Using these relations and considering some corrections on the transition line, 
we study the critical behaviors of the spin chain (\ref{hamiltonian}) 
numerically. 

\subsection{Twisted boundary method to determine the $c=1$ critical line}
\label{c1}
In this subsection, we consider the numerical method to determine the critical 
point. Firstly let us investigate the symmetry of the system. 
For the periodic boundary condition, 
the Hamiltonian (\ref{hamiltonian}) is invariant 
under two-site translation 
$\mbox{\boldmath$S$}_{j}\rightarrow \mbox{\boldmath$S$}_{j+2}$, 
space inversion 
$\mbox{\boldmath$S$}_{j}\rightarrow \mbox{\boldmath$S$}_{N-j+1}$, 
and spin reversal $S_{j}^{z}\rightarrow -S_{j}^{z}$. 
Eigenstates have corresponding quantum numbers of 
the wave number ($q=-\pi-4\pi/N, -\pi -8\pi/N,\cdots, \pi$), 
of the parity ($P=\pm1$), and of the spin reversal ($T=\pm 1$). 
In a variational way, the ground state of the Haldane phase 
can be described by Schwinger bosons as\cite{Arovas}
\begin{equation}
  (a^{\dagger}_{N}b^{\dagger}_{1}-b^{\dagger}_{N}a^{\dagger}_{1})
  \prod_{j=1}^{N-1}
  (a^{\dagger}_{j}b^{\dagger}_{j+1}-b^{\dagger}_{j}a^{\dagger}_{j+1})
  |0\rangle,
\label{halstate}
\end{equation}
while the ground state of the dimer state can be written as 
\begin{equation}
  \prod_{j=1}^{N/2}
    (a^{\dagger}_{2j-1}b^{\dagger}_{2j}-b^{\dagger}_{2j-1}a^{\dagger}_{2j})^{2}
  |0\rangle, \hspace{5mm} (\delta>0)
\label{dimstate}
\end{equation}
or
\[
  (a^{\dagger}_{N}b^{\dagger}_{1}-b^{\dagger}_{N}a^{\dagger}_{1})^{2}
  \prod_{j=1}^{N/2-1}
    (a^{\dagger}_{2j}b^{\dagger}_{2j+1}-b^{\dagger}_{2j}a^{\dagger}_{2j+1})^{2}
  |0\rangle, \hspace{5mm} (\delta<0)
\]
where $a^\dagger_{j}$ and $b^{\dagger}$ are Schwinger bosons to create the 
$S^{z}=+1/2$ and $S^{z}=-1/2$, and $|0\rangle$ is the vacuum state for these 
bosons. 
In this boson representation, space inversion is 
$P:a^{\dagger}_{j}\rightarrow a^{\dagger}_{L-j+1}$, 
$b^{\dagger}_{j}\rightarrow b^{\dagger}_{L-j+1}$, and spin reversal is 
$T:a^{\dagger}_{j}\leftrightarrow b^{\dagger}_{j}$. 
For the periodic boundary condition, the ground states in both the Haldane 
and the dimer phases are singlet with $q=0$, $P=1$, and $T=1$.

To determine the transition point numerically between the Haldane 
and the dimer phases, 
we use a trick in refs. \cite{KN97b,K97,KN97a} with 
the twisted boundary condition (TBC)
\begin{equation}
S^{x}_{N+1}\pm iS^{y}_{N+1}=-(S^{x}_{1}\pm iS^{y}_{1}),\hspace{5mm} 
S^{z}_{N+1}=S^{z}_{1}.
\label{twst}
\end{equation} 
This boundary condition distinguishes the above two states, that is, 
for the Haldane phase
\begin{equation}
  (a^{\dagger}_{N}b^{\dagger}_{1}+b^{\dagger}_{N}a^{\dagger}_{1})
  \prod_{j=1}^{N-1}
  (a^{\dagger}_{j}b^{\dagger}_{j+1}-b^{\dagger}_{j}a^{\dagger}_{j+1})
  |0\rangle,
\end{equation}
which has $P=T=-1$, and for the dimer state ($\delta>0$)
\begin{equation}
  \prod_{j=1}^{N/2}
    (a^{\dagger}_{2j-1}b^{\dagger}_{2j}-b^{\dagger}_{2j-1}a^{\dagger}_{2j})^{2}
  |0\rangle,
\end{equation}
or ($\delta<0$)
\begin{equation}
  (a^{\dagger}_{N}b^{\dagger}_{1}+b^{\dagger}_{N}a^{\dagger}_{1})^{2}
  \prod_{j=1}^{N/2-1}
    (a^{\dagger}_{2j}b^{\dagger}_{2j+1}-b^{\dagger}_{2j}a^{\dagger}_{2j+1})^{2}
  |0\rangle,
\end{equation}
which have $P=T=1$. 
These states have different eigenvalues for the space inversion ($P$)
and the spin reversal ($T$). 
Thus we expect that 
we can determine the phase boundary numerically 
by the level crossing with this boundary condition. 

To verify that the level crossing indicates a phase transition, 
let us consider the above behavior under TBC in another point of view. 
Affleck and Haldane\cite{Affleck87} argued that 
the critical properties of the considering transition point 
is of the $k=1$ SU(2) WZW model. 
In this context, the phase transition is described by the following model, 
\begin{eqnarray}
  H &=& \frac{1}{2\pi}\int dx\left[ 
  K(\pi \Pi)^{2}+\frac{1}{K}\left(\frac{\partial\phi}{\partial x}\right)^{2} 
  \right] \nonumber \\
  && -\frac{y_{1}}{2\pi}\int dx :\sin\sqrt{2}\phi :
  -\frac{y_{0}}{2\pi}\int dx :\cos 2\sqrt{2}\phi:, 
\label{sine}
\end{eqnarray}
where $\Pi$ is the momentum density conjugate to $\phi$, 
and to assure the SU(2) symmetry the couplings $K$ and $y_{0}$ 
are not independent ($2K-2 = y_{0}$, near $K=1$).\cite{Cardy87} 
The dual field $\tilde{\phi}$ is defined as 
$\partial_{x}\tilde{\phi}(x) = \pi\Pi$. 
Since under TBC the system loses the SU(2) 
symmetry, this model is useful for seeing 
the effect of the boundary condition. 
In this model, the operator $\sqrt{2}:\sin\sqrt{2}\phi:$ in the second term 
is a primary field of the $k=1$ SU(2) WZW theory with 
the scaling dimension $x=1/2$ and with quantum numbers 
$s=1/2$, $\bar{s}=1/2$, $S=0$. 
This operator is relevant and a second order phase transition occurs at 
$y_{1}=0$. 
The renormalization group equation is given by 
\begin{eqnarray}
  \frac{d y_{1}(L)}{d\ln L} &=& \frac{3}{2}y_{1}(L)
  -\frac{3}{4}y_{1}(L)y_{0}(L), 
  \nonumber \\
  \frac{d y_{0}(L)}{d\ln L} &=& -\frac{1}{4}(y_{1}(L))^{2}-(y_{0}(L))^{2}. 
\label{rgg}
\end{eqnarray}

It has been known that for a primary field 
$:\exp(i m\phi+i n\tilde{\phi}):$, where integers $m$ and $n$ are magnetic 
and electric charges in the Coulomb gas, 
the effect of the twisted boundary 
condition\cite{Blote,Alcaraz87,Destri,Fukui} 
is to shift the magnetic charge 
$m$ as $m\rightarrow m+1/2$, and $n\rightarrow n$. 
($m$ and $n$ are related to the U(1) charges 
$\alpha$, $\bar{\alpha}$ as 
$\alpha-\bar{\alpha}=\sqrt{2}m$, and $\alpha+\bar{\alpha}=\sqrt{2}n$.) 
Then the operators with the lowest scaling dimensions are 
$:\exp(\pm i\phi/\sqrt{2}):$, and the second term of eq. (\ref{sine}) 
hybridizes these operators into 
${\cal{O}}_{\pm}=\sqrt{2}:\cos(\phi/\sqrt{2}\pm\pi/4):$. 
From the first order perturbation calculation, size dependence of the scaling 
dimensions of these half magnetic charge operators 
is given by\cite{K97}
\begin{eqnarray}
  x_{+}(L) &=& \frac{1}{8}+\frac{y_{1}(L)}{2} + \frac{y_{0}(L)}{16},
\nonumber \\
  x_{-}(L) &=& \frac{1}{8}-\frac{y_{1}(L)}{2} + \frac{y_{0}(L)}{16},
\label{szdph}
\end{eqnarray}
respectively.  $y_{1}(L)$ and $y_{0}(L)$ are the solution of eq. (\ref{rgg}). 
Scaling dimensions are related to the excitation energy 
of the finite size system as eq. (\ref{dimension}), 
thus low lying two levels with the twisted boundary condition (\ref{twst}) 
cross linearly at the transition point $y_{1}=0$. 
At $y_{1}=0$, $y_{0}(L)$ term in eq. (\ref{szdph}) gives a logarithmic 
dependence of the system size ($y_{0}(L) = y_{0}/(y_{0}\ln L +1)$). 
Hence we verified that for the phase transition of the $k=1$ SU(2) WZW type, 
we must observe a level crossing of two low lying energies under 
the twisted boundary conditions. 

This level crossing is similar to the findings 
of F\'ath and S\'olyom.\cite{Fath} 
For $\delta=0$ they studied 
the model (\ref{hamiltonian}) numerically 
with the general twisted boundary condition 
$S^{\pm}_{N+1} = e^{\pm i\Theta}S^{\pm}_{1}$, $S^{z}_{N+1}=S^{z}_{1}$, 
and they found that for $\Theta=\pi$ two levels 
(in their notation, $a$ and $d_{0}$, see Fig. 4 in their paper) 
cross at the TB point. 
But in our method, we see different level crossing from them 
(in ref\cite{Fath}, we see the crossing $a$ and $c_{0}$ (or $b$)). 
Level crossing in ref. \cite{Fath} can be related to a $c=3/2$ CFT, 
and we can not apply it directly to the $c=1$ case. 
To see this, we show the $c=3/2$ behavior in Appendix. 

\subsection{Numerical results}
Here we show the numerical results of $N=8,10,12,14,16$ systems with the 
exact diagonalization. 
We firstly determine the phase boundary between the Haldane and the dimer 
phases numerically (discussed in subsection \ref{c1}), 
and next see the critical behaviors of the system on the $c=1$ line 
(discussed in section II). 

Figure 3 shows the two low-lying energies 
of the subspace $\sum S^{z}=0$ with the twisted boundary condition 
for $N=12$ and $\delta=0.20$. 
These two states have different eigenvalues of 
the space inversion and the spin reversal operators. 
To compare with the periodic boundary condition case, 
we subtract the ground state energy $E_{g}(L)$. 
For $\theta < \theta_{c}$, the lower state has $P=T=1$, while 
for $\theta > \theta_{c}$ the lower state has $P=T=-1$. 
This is the expected behavior as is described in \ref{c1}. 

To see the size dependence we show the crossing points of $N=8,12,16$ 
systems in Fig. 4. 
For large $\delta$, the size dependence of the crossing points is small. 
But near the TB point ($\theta=-\pi/4$, $\delta=0$) where $c=3/2$, 
the convergence is very slow. 
In Fig. 5, we show the size dependence for $\delta=0$, 
and extrapolate data with polynomials of $1/N$. 
The extrapolated value is $\theta=-0.232\pi$ which deviates 
from the ideal value. 

To obtain the phase diagram, we extrapolate the data 
in the region $0.03 < \delta<0.8$. 
To calculate the scaling dimension and the conformal anomaly, 
we interpolate the extrapolated points with cubic spline 
in the region $-\pi/4<\theta<\mbox{arctan}(1/3)$, $0<\delta<1$. 
To compare with the behavior (\ref{eq:line}) of the transition line, 
we show the line $\delta=(\theta/\pi+0.25)^{13/8}$ in Fig. 4(b), 
but in this interpolation, we can not detect 
the precise exponent $13/8$ of the critical line (\ref{eq:line}). 

Near the another limit $(\theta,\delta)=(\mbox{arctan}(1/3),\pm 1)$, 
there is the following variational calculation result.\cite{Totsuka} 
Using the states (\ref{halstate}) and (\ref{dimstate}), 
the transition line between the Haldane and the dimer phases is given by 
\begin{equation}
  \delta = \mp\frac{2\tan\theta+1}{4\tan\theta-3}. 
\label{variation}
\end{equation}
Of course, this line is only valid near the point 
$(\theta,\delta)=(\mbox{arctan}(1/3),\pm 1)$, where states 
(\ref{halstate}) and (\ref{dimstate}) are exact degenerate ground states. 
In Fig. 4(a), we also show this line. 

We calculate the conformal anomaly on the transition line 
by the following equation
\begin{equation}
  \frac{E_{g}(L)}{L} = \epsilon_{g}-\frac{\pi vc}{6L^{2}}
  + b\frac{1}{L^{4}}, \hspace{5mm}(L=N/2). 
\end{equation}
The sound velocity $v$ is calculated from the lowest energy 
with the momentum $q=4\pi/N$ and the total spin 1
which corresponds to the SU(2) current. 
We determine the coefficients $\epsilon_{g}$, $c$, and $b$, using 
$N$, $N+2$, and $N+4$ site systems. 
Figure 6 shows the obtained values of $c(N)$. 
Reflecting the $c$-theorem, the conformal anomaly decreases 
from $c=3/2$ to $c=1$, and 
increasing the system size the crossover region of this behavior decreases. 
We can see the asymptotic behavior of eq. (\ref{aanmly}) 
in Fig. 6(b). 
There exists also logarithmic correction in $c(L)$ (ground state energy) 
as \cite{Ludwig,Affleck89,Cardy86} 
\[
  c(L) = \frac{3k}{2+k} + \frac{3k^{2}}{8}(\ln L)^{-3},
\]
which may be small in our numerical results (Fig. 6).

Using eq. (\ref{dimension}), we also calculate the scaling dimensions 
corresponding to the primary fields of the $k=2$ SU(2) WZW model 
in Figs. 7 and 8. 
These figures show the values of $N=16$ systems with $q=0$, and the total spin 
$S=0$, $1$, and $2$ excitations. 
These scaling dimensions split due to the marginal corrections 
which depend on the spin of the excitation,\cite{Ziman,Affleck89} 
\begin{eqnarray}
  \frac{L\Delta E_{s,\bar{s}}^{S,S^{z}}(L)}{2\pi v}
  &=& \frac{s(s+1)}{k+2}+\frac{\bar{s}(\bar{s}+1)}{k+2} \\
  &&-\frac{1}{2}\frac{[S(S+1) - s(s+1)- \bar{s}(\bar{s}+1)]}{\ln L}
  \nonumber
\end{eqnarray}
In the same figure, we also show averaged scaling dimensions 
with appropriate weights to eliminate the logarithmic correction.\cite{Ziman} 
Although we do not extrapolate the data, the averaged scaling dimensions 
are very close to the ideal value $x=3/8$ and $x=1$ at the TB point. 
We see that in the $c=1$ region the scaling dimensions 
with $s=\bar{s} = 1/2$ and $1$ 
are $x_{1/2,1/2}=1/2$ and $x_{1,1}=2$ respectively. 
Especially the operators with $s=\bar{s} = 1$ move to marginal ones. 
Comparing $x_{1/2,1/2}$ and $x_{1,1}$, we see that 
$x_{1,1}$ converges slowly to the expected value $x=2$, 
which may come from the finite size correction 
of the descendent fields of the identity 
operator $L_{4}{\cal{I}}$, $\bar{L}_{4}{\cal{I}}$, 
and $L_{2}\bar{L}_{2}{\cal{I}}$.\cite{Cardy,Reinicke} 

\section{Summary and Discussion}
In this paper, we studied the $S=1$ spin chains with bilinear, biquadratic, 
plus bond-alternating interactions, to see the crossover behavior 
from the Takhtajan-Babujian point ($\theta=-\pi/4$, $\delta=0$, $c=3/2$) 
to the $c=1$ critical line. 
This line separates the Haldane and the dimer phases. 
Based on the $k=2$ SU(2) WZW model, we argued the asymptotic behavior 
near the Takhtajan-Babujian point. 
To determine the transition point numerically, 
we used the twisted boundary condition on the basis of a $c=1$ model. 
This method directly see the lower energy between two different states, 
which converge to the ground state energy 
of each phases in the thermodynamic limit. 

With semiclassical and topological arguments,  
Affleck and Haldane\cite{Affleck87} 
predicted the multi-critical structure for spin models with bond-alternation. 
There should be $2S$ $k=1$ SU(2) WZW critical line and 
a first order phase transition 
line and these lines meet at the spin $S$ Takhtajan-Babujian point. 
In the arbitrary $S$ TB model, primary fields with the conformal spin 
$h-\bar{h}=0$ are $\Phi_{s,\bar{s}}$, $s=\bar{s}=0,1/2,1,\cdots, 2S$, 
in which operators with the half integer $s=\bar{s}$ 
have the momentum with $q=\pi$, and others have $q=0$. 
For the one-site translational invariant system, operators 
with half integer $s=\bar{s}$ cannot be a relevant operators 
in the Hamiltonian, because these operators violate 
the translational invariance. 
Bond-alternation makes these operators into relevant one. 
Among these operators, $\Phi_{1/2,1/2}^{0,0}$ and $\Phi_{1,1}^{0,0}$ are 
important for the crossover behavior.\cite{Affleck87} 
$\Phi_{1/2,1/2}^{S,S^{z}}$ shifts to the primary fields 
in the $k=1$ WZW theory, and $\Phi_{1,1}^{S,S^{z}}$ moves to the marginal 
operators $J^{a}\bar{J}^{b}$. 

We saw the behaviors of the conformal anomaly 
and the scaling dimensions of some operators 
which are the primary fields for the 
$k=2$ SU(2) WZW theory. 
This behavior may also be valid for bond-alternating higher spin cases, 
especially as the generalization of eq. (\ref{eq:line}), 
the transition line behaves as 
$\delta\sim (g_{2})^{(8S+5)/8S}$ for $S>1/2$  models. 

\acknowledgments
This work is partly supported by a Grant-in-Aid for Scientific 
Research (c) No. 09740308 from the Ministry of Education, Science, 
Sports, and Culture, Japan.
A. K. is supported by JSPS Research Fellowships for Young Scientists.
The numerical calculation was performed 
using the facilities of the Super Computer Center, 
Institute for Solid State Physics, University of Tokyo.

\appendix
\section{Twisted boundary conditions 
on the $\delta=0$ line near the TB point}
F\'ath and S\'olyom\cite{Fath} found that with the twisted boundary condition, 
\begin{equation}
  S^{x}_{N+1}\pm iS^{y}_{N+1} = e^{\pm i\Theta} S^{x}_{1}\pm iS^{y}_{1}, 
  \hspace{5mm} S^{z}_{N+1} = S^{z}_{1}
\label{twist}
\end{equation}
a level-crossing occurs for $\Theta=\pi$ at the TB point 
on the line $\delta=0$. 
Here we show this behavior, considering the finite size correction 
(up to the leading logarithmic one). 

For the twisted boundary condition, the partial bosonized model is useful. 
The critical properties of two independent 2D Ising models can be described by 
the Dirac fermions. This system can be identified with a Gaussian model. 
Bosonizing (\ref{WZW}) and (\ref{Majorana}) 
about the fields $\psi_{1}$, $\bar{\psi}_{1}$, 
$\psi_{2}$, and $\bar{\psi}_{2}$, we have the following action 
\begin{equation}
  {\cal{S}} = {\cal{S}}_{0} + {\cal{S}}_{I},
\end{equation}
where the free part is given by
\begin{equation}
  {\cal{S}}_{0} = \int \frac{d^{2}r}{2\pi}\frac{1}{2}(\partial_{\mu}\phi)^{2}
  + \int \frac{d^{2}r}{2\pi}\left( \psi_{3}\bar{\partial}\psi_{3} + 
  \bar{\psi}_{3}\partial\bar{\psi}_{3} \right),
\label{action0}
\end{equation}
where $d^{2}r=d\tau dx$, and we rescaled the Fermi field 
as $\psi\rightarrow \psi^{'}=\sqrt{2\pi}\psi$. 
${\cal{S}}_{I}$ is the interaction part, 
\begin{equation}
  {\cal{S}}_{I} = -g_{2}\int\frac{d^{2}r}{2\pi}\Phi^{0,0}_{1,1}(r)
  -g_{0}\int\frac{d^{2}r}{2\pi}\frac{1}{\sqrt{3}}
  \mbox{\boldmath$J$}(r)\cdot\bar{\mbox{\boldmath$J$}}(r). 
\end{equation}
A second order phase transition occurs at $g_{2}=0$. 
With the correspondence for the energy density 
\begin{equation}
 \frac{1}{\sqrt{2}}(i\psi_{1}\bar{\psi}_{1}+i\psi_{2}\bar{\psi}_{2}) 
  = \sqrt{2}:\cos\sqrt{2}\phi:,
\end{equation}
and the polarization operator of the Ashkin-Teller model
\begin{equation}
  \sigma_{1}\sigma_{2} = \sqrt{2}:\sin\left(\frac{\sqrt{2}}{2}\phi \right):,
\end{equation}
we have the following operators with $S^{z}_{0} = 0$
\begin{eqnarray}
  \Phi_{1/2,1/2}^{0,0} &=& \sqrt{2}:\sin\frac{\sqrt{2}}{2}\phi :
  \sigma_{3}, \nonumber \\
  \Phi_{1/2,1/2}^{1,0} &=& \sqrt{2}:\cos\frac{\sqrt{2}}{2}\phi :
  \sigma_{3}, \nonumber \\
  \Phi_{1,1}^{0,0} &=& \frac{1}{\sqrt{3}}\left( -2:\cos\sqrt{2}\phi :
  + i\psi_{3}\bar{\psi}_{3}\right), 
\label{bosonop} \\
  \Phi_{1,1}^{1,0} &=& \sqrt{2}:\sin\sqrt{2}\phi :,  \nonumber \\
  \Phi_{1,1}^{2,0} &=& \frac{1}{\sqrt{6}}\left( 2i\psi_{3}\bar{\psi}_{3}
  +2:\cos\sqrt{2}\phi :\right). \nonumber
\end{eqnarray}
In ref. \cite{Fath}, $\Phi_{1/2,1/2}^{0,0}$ corresponds to $b$, 
$\Phi_{1/2,1/2}^{1,0}$ to $c_{0}$, and 
$\Phi_{1,1}^{2,0}$ to $d_{0}$. 

The third component of the SU(2) current can be written as
\begin{equation}
  J^{3} = i\partial\phi,
  \hspace{5mm} \bar{J}^{3} = -i\bar{\partial}\phi,
\end{equation}
so that $J^{3}$ and $\bar{J}^{3}$ are U(1) currents 
of the bosonic Gaussian part of eq. (\ref{action0}), 
and the boundary condition (\ref{twist}) 
does not affect to the fields $\psi_{3}$ and $\bar{\psi}_{3}$. 
For $g_{2}=g_{0}=0$, the twisted boundary condition 
(\ref{twist}) changes the operators $\Phi_{1,1}^{S^{z},0}$ 
in eq. (\ref{bosonop}) to the following operators 
\begin{eqnarray}
{\cal{O}}_{1}(\Phi) &=& :\exp\left[i\sqrt{2}(1+\frac{\Theta}{2\pi})\phi
    \right]:,   \nonumber \\
{\cal{O}}_{2}(\Phi) &=& :\exp\left[i\sqrt{2}(-1+\frac{\Theta}{2\pi})\phi
    \right]:, 
  \\
{\cal{O}}_{3} &=& i\psi_{3}\bar{\psi}_{3},
  \nonumber 
\end{eqnarray}
and the corresponding excitation energy for the finite size system changes to 
\begin{eqnarray}
  E_{1}(\Theta)-E_{g}(0) &=& 
    \frac{2\pi v}{L}\left(1+\frac{\Theta}{2\pi}\right)^{2},
\nonumber \\
  E_{2}(\Theta)-E_{g}(0) &=& 
    \frac{2\pi v}{L}\left(-1+\frac{\Theta}{2\pi}\right)^{2},
\nonumber \\
  E_{3}(\Theta)-E_{g}(0) &=& \frac{2\pi v}{L}\times 1,
\nonumber
\end{eqnarray}
where $E_{g}(0)$ is the ground state energy 
with the periodic boundary condition. 
Thus excitation levels split by the twisted boundary condition. 
The identity operator ${\cal{I}}$ 
(corresponding to the ground state, in ref. \cite{Fath} to $a$) 
also changes to the operator 
\begin{equation}
  {\cal{I}}(\Theta) = :\exp\left[i\sqrt{2}\frac{\Theta}{2\pi}\phi\right]:, 
\end{equation}
and correspondingly the energy is 
\begin{equation}
  E_{g}(\Theta)-E_{g}(0) 
  = \frac{2\pi}{L}\left(\frac{\Theta}{2\pi}\right)^{2}.
\end{equation}

So far we have considered the free field theory $g_{2}=g_{0}=0$. 
From above equations, we see that $E_{g}(\Theta)$ and $E_{2}(\Theta)$ 
are equal at $\Theta=\pi$, and 
for non-zero coupling case, states $|{\cal{O}}_{2}(\Theta)\rangle$ 
and $|{\cal{I}}(\Theta)\rangle$ are hybridized 
by these couplings at $\Phi=\pi$, 
to the operators
\begin{eqnarray}
  {\cal{O}}_{e} &=& \frac{1}{\sqrt{2}}:\cos\frac{\sqrt{2}}{2}\phi:
  (= \mu_{1}\mu_{2}),  \nonumber \\
  {\cal{O}}_{o} &=& \frac{1}{\sqrt{2}}:\sin\frac{\sqrt{2}}{2}\phi:
  (= \sigma_{1}\sigma_{2}). 
\label{polar}
\end{eqnarray}
From OPEs
\begin{eqnarray}
  \Phi^{0,0}_{1,1}(z,\bar{z}){\cal{O}}_{e}(w,\bar{w})
  &=& -\frac{1}{\sqrt{3}}\frac{1}{|z-w|^{2}}{\cal{O}}_{e}(w,\bar{w})
  +\cdots, \nonumber \\
  \Phi^{0,0}_{1,1}(z,\bar{z}){\cal{O}}_{o}(w,\bar{w})
  &=& \frac{1}{\sqrt{3}}\frac{1}{|z-w|^{2}}{\cal{O}}_{o}(w,\bar{w})
  +\cdots, \nonumber \\
  \frac{1}{\sqrt{3}}\mbox{\boldmath$J$}(z)\cdot
  \bar{\mbox{\boldmath$J$}}(\bar{z}){\cal{O}}_{e,o}(w,\bar{w})
  &=& -\frac{1}{\sqrt{3}}\frac{1}{|z-w|^{4}}{\cal{O}}_{e}(w,\bar{w})
  +\cdots, \nonumber
\end{eqnarray}
the first order perturbation calculation of the energy is given by 
\begin{eqnarray}
  \lefteqn{E_{e}(\pi)-E_{g}(0)} \nonumber \\
   &=& \frac{2\pi}{L}\left[
  \frac{1}{4} + \frac{1}{\sqrt{3}}g_{2}(L)
  +\frac{1}{\sqrt{3}}g_{0}(L)  \right],
\nonumber \\
  \lefteqn{E_{o}(\pi)-E_{g}(0)}  
\label{eq:ap} \\
  &=& \frac{2\pi}{L}\left[
  \frac{1}{4} - \frac{1}{\sqrt{3}}g_{2}(L)
  +\frac{1}{\sqrt{3}}g_{0}(L) \right],
\nonumber  
\end{eqnarray}
respectively. 
$g_{2}(L)$ and $g_{0}(L)$ are the solution of 
the following renormalization group equations, 
\begin{eqnarray}
  \frac{d g_{2}(L)}{d\ln L} &=& g_{2}(L) - \frac{2\sqrt{3}}{3}g_{2}(L)g_{0}(L),
  \nonumber \\ 
  \frac{d g_{0}(L)}{d\ln L} &=& -\frac{\sqrt{3}}{3}(g_{2}(L))^{2}
  -\frac{\sqrt{3}}{3}(g_{0}(L))^{2}. 
\end{eqnarray}
In eq. (\ref{eq:ap}) $g_{0}(L)$ gives the logarithmic correction 
at the TB point. 
Hence the energy levels $a$ and $d_{0}$ with $\Phi=\pi$, that is, 
$E_{e}(\pi)$ and $E_{o}(\pi)$ cross linearly at the TB point ($g_{2}=0$).

\begin{figure}
\begin{center}
\leavevmode \epsfxsize=3.2in \epsfbox{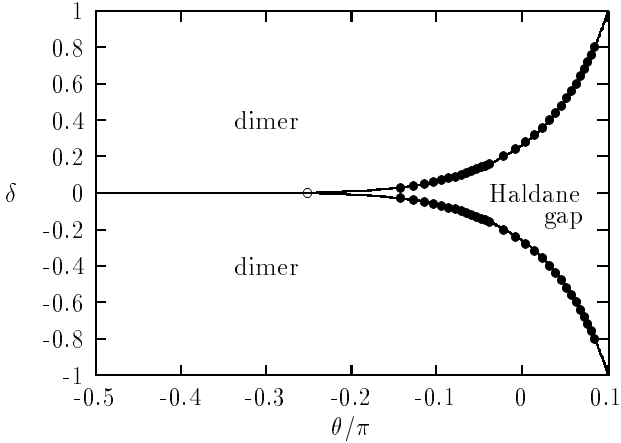}
\end{center}
\caption{Phase diagram of the $S=1$ spin chains (\ref{hamiltonian}). 
TB point is $\theta=-\pi/4$ and $\delta=0$ ($\circ$), 
which is the multi-critical point. 
Lines in the region $-\pi/4<\theta<\mbox{arctan}(1/3)$ are 
$c=1$ critical lines. 
$\bullet$ is the extrapolated numerical data 
for $N=8,10,12,14,16$ systems (section III). 
The line $\delta=0$ $\theta<-\theta/4$ is 
a first order phase transition line.\cite{Affleck87,Solyom} }
\end{figure}

\begin{figure}
\begin{center}
\leavevmode\epsfxsize=3.2in \epsfbox{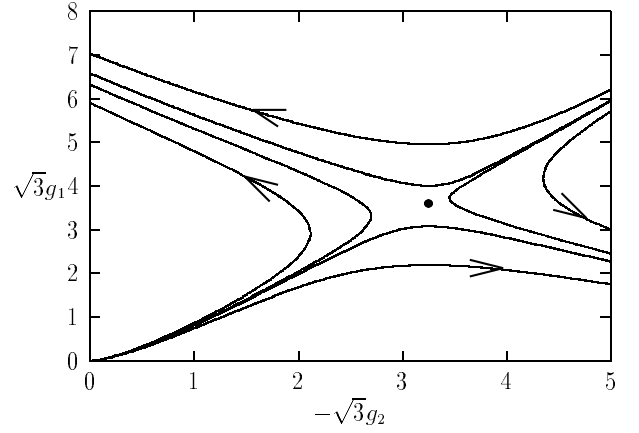}
\end{center}
\caption{Renormalization group flow.}
\end{figure}

\begin{figure}
\begin{center}
\leavevmode\epsfxsize=2.8in \epsfbox{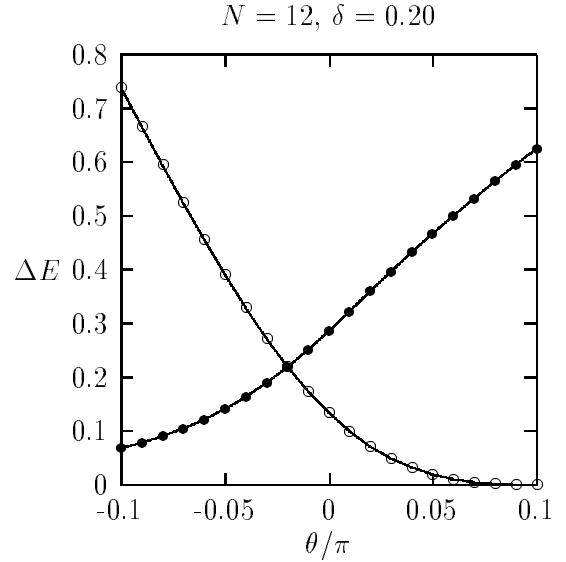}
\end{center}
\caption{Level crossing in the system 
with the twisted boundary conditions.
$\bullet$ is the lowest $P=T=1$ energy $E_{e}-E_{g}$, 
and $\circ$ is the lowest 
$P=T=-1$ energy $E_{o}-E_{g}$ for the $N=12$ systems. 
We subtract the ground state energy with the periodic boundary condition.}
\end{figure}

\begin{figure}
\begin{center}
\leavevmode\epsfxsize=3.2in \epsfbox{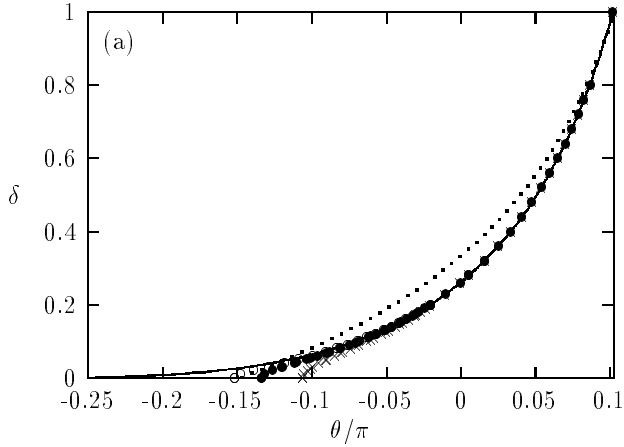}
\vspace{5mm}

\leavevmode\epsfxsize=3.2in \epsfbox{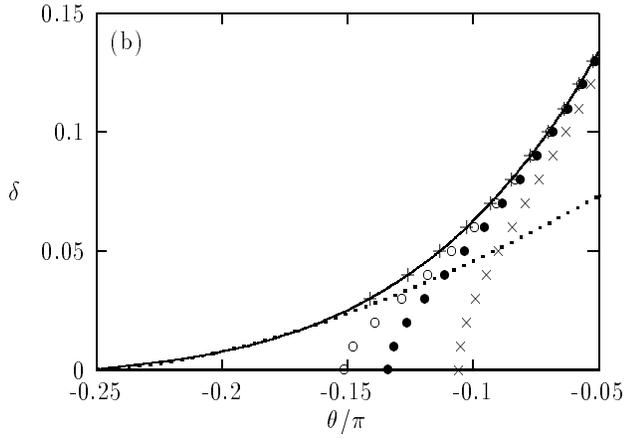}
\end{center}
\caption{Crossing points for $N=8$ ($\times$), $N=12$ ($\bullet$), 
and $N=16$ ($\circ$) systems. 
$+$'s are the extrapolated data, 
and the solid line is the interpolated one of it. Dotted lines are (a) 
eq. (\ref{variation}) and (b) $\delta=(\theta/\pi+0.25)^{13/8}$.}
\end{figure}

\begin{figure}
\begin{center}
\leavevmode\epsfxsize=3.2in \epsfbox{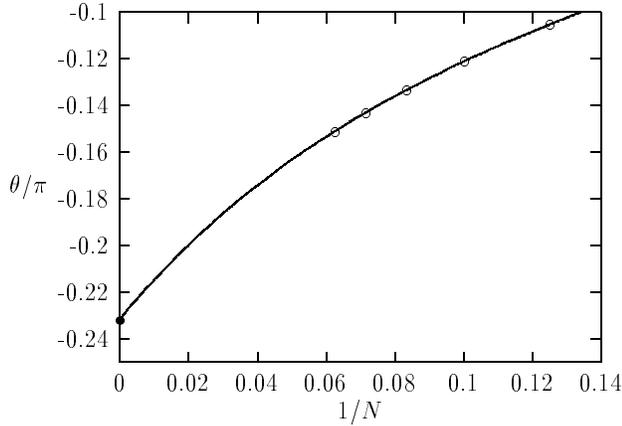}
\end{center}
\caption{Size dependence of the crossing point for $\delta=0$ as 
a function of $1/N$. The extrapolated value ($\bullet$) 
is $\theta = -0.232\pi$.}
\end{figure}

\begin{figure}
\begin{center}
\leavevmode\epsfxsize=3.2in \epsfbox{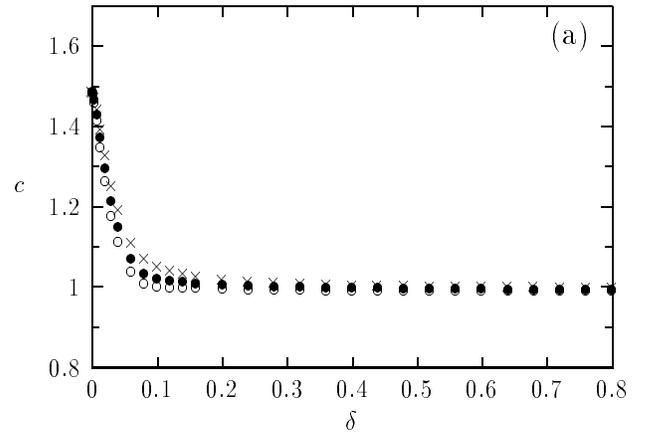}
\vspace{5mm}

\leavevmode\epsfxsize=3.2in \epsfbox{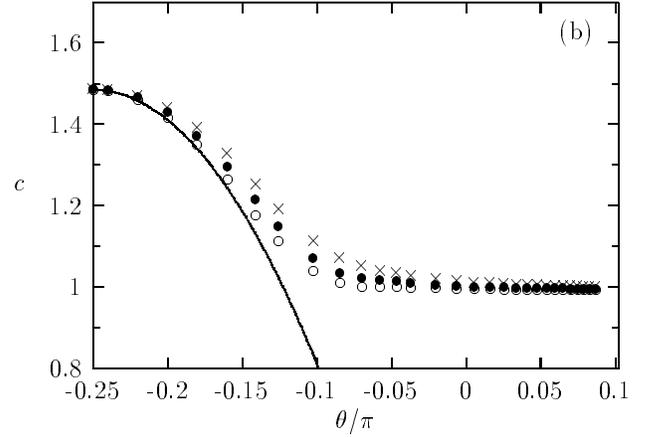}
\end{center}
\caption{Conformal anomaly $c(N)$ on the $c=1$ critical line 
for $N=8$ ($\times$), $N=10$ ($\bullet$), 
and $N=12$ ($\circ$) systems, as a function of (a) $\delta_{c}$ 
and (b) $\theta_{c}$. 
In (b), we also show the line $c_{TB}-\alpha(\theta+0.25\pi)^{2}$, 
in which numerical number $c_{TB}$ and $\alpha$ are determined by data 
at $\theta=-0.25\pi$ and $-0.24\pi$ for $c(N=12)$.}
\end{figure}

\begin{figure}
\begin{center}
\leavevmode\epsfxsize=3.2in \epsfbox{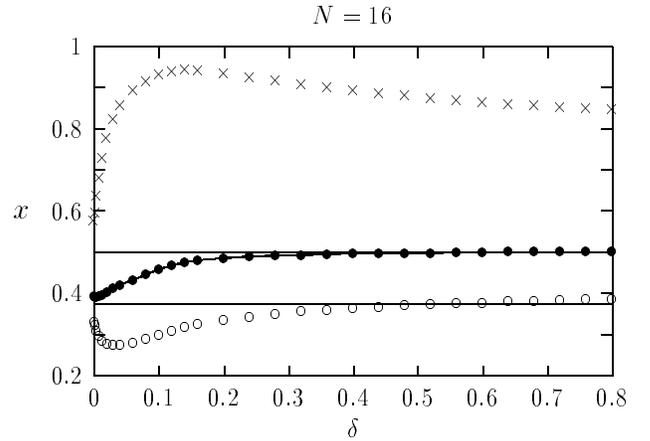}
\end{center}
\caption{Scaling dimensions ($x_{1/2,1/2}^{S}$) of $\Phi_{1/2,1/2}^{S,S^{z}}$ 
obtained by eq. (\ref{dimension}) for the $N=16$ system. 
$\times$'s are of $S=0$ and $\circ$'s are of $S=1$. 
$\bullet$'s are the averaged value $(x_{1/2,1/2}^{0}+3x_{1/2,1/2}^{1})/4$. 
Horizontal lines are $x=3/8$ and $x=1/2$.}
\end{figure}

\begin{figure}
\begin{center}
\leavevmode\epsfxsize=3.2in \epsfbox{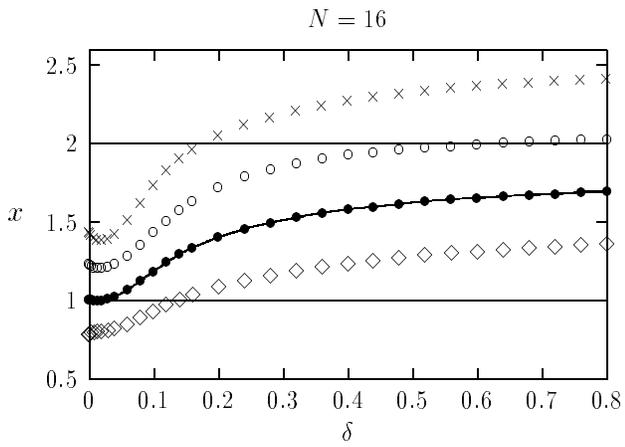}
\end{center}
\caption{Scaling dimensions of $\Phi_{1,1}^{S,S^{z}}$ 
obtained by eq. (\ref{dimension}) for the $N=16$ system. 
$\times$'s are of $S=0$, $\circ$'s of $S=1$, and $\Diamond$'s of $S=2$. 
$\bullet$'s are the averaged value 
$(x_{1,1}^{0}+3x_{1,1}^{1}+5x_{1,1}^{2})/9$. 
Horizontal lines are $x=1$ and $x=2$.}
\end{figure}

\end{document}